\documentclass[twocolumn]{aastex7}
\begin{document}

% for float placement:
\renewcommand{\topfraction}{1.0}
\renewcommand{\bottomfraction}{1.0}
\renewcommand{\textfraction}{0.0}

\newcommand{\kms}{km~s$^{-1}$\,}
\newcommand{\masyr}{mas~yr$^{-1}$\,}
\newcommand{\msun}{$M_\odot$\,}

\title{Spectroscopic Orbits of Subsystems in  Multiple
  Stars. XI}

\author[orcid=0000-0002-2084-0782]{Andrei Tokovinin}
\affiliation{Cerro Tololo Inter-American Observatory, NSF's NOIRLab
Casilla 603, La Serena, Chile}
\email{andrei.tokovinin@noirlab.edu}

\begin{abstract}
Adding to the large radial  velocity survey of nearby solar-type stars
(summary in Tokovinin, 2023a), spectroscopic orbits are determined for
four  hierarchical systems:  HIP  49442 (inner  and  outer periods  of
164.55 d  and 34 yr,  respectively), HIP 55691  (2.4 and 415  yr), HIP
61465 (86.8  d), and HIP 78662C  (0.82 d).  Each system  is discussed
individually.  Seven  Gaia orbits  of low-mass  dwarfs, each  with two
additional  resolved  (interferometric   and  wide)  companions,  i.e.
potential quadruples, are tested by monitoring radial velocities; five
orbits are confirmed  and two are refuted.  Five of  these systems are
quadruples  of 3+1  hierarchy, one  is quintuple,  and one  is triple.
Strengths and limitations of the Gaia data on multiple systems and the
need of complementary observations are highlighted.
\end{abstract}


%\keywords{\uat{Galaxies}{573} --- \uat{Cosmology}{343}

%% https://astrothesaurus.org

%\maketitle

%---------------------------------------------------------
\section{Introduction}
\label{sec:intro}

Hierarchical  systems  of  three  or   more  stars  are  an  important
constituent of the stellar population.  Statistical knowledge of their
architecture  (distributions of  periods, eccentricities,  masses, and
multual  inclinations)  is informative  both  for  the star  formation
theory and for the population synthesis of stellar evolution. Owing to
the  huge range  of  periods  and separations,  coverage  of the  full
parameter space requires combination  of several observing techniques,
and at  present it  is still very  incomplete. This  work contributes
spectroscopic orbits of several nearby hierarchies.

The Gaia  mission \citep{Gaia1,Gaia3} provides precise  astrometry and
other information uniformly for the  whole sky.  Within 100\,pc, there
are about 8000 pairs of  Gaia sources with separations above $\sim$100
au and  indications of  an inner subsystem  (an elevated  reduced unit
weight error, RUWE, or multi-peak  transits) in one or both components
\citep{Tok2023}.  However, Gaia informs us  on the parameters of these
subsystems  only for  a  small  subset of  stars  with astrometric  or
spectroscopic  orbits  (inner  periods   below  3  yr).   Furthermore,
hierarchies with outer separations under  100 au (the most interesting
ones)  are  revealed  by  Gaia only  exceptionally  by  cross-matching
different  multiplicity  indicators  \citep{Czavalinga2023,Bashi2024}.
Parallaxes  of many  pairs with  separations under  1\arcsec ~are  not
provided in  the Gaia data  release 3  (GDR3), and those  binaries and
hierarchies are  missing from the  Gaia catalog of nearby  stars, GCNS
\citep{GCNS}.  Even within 20 pc,  a significant fraction of stars and
brown dwarfs lack Gaia astrometry, their multiplicity being one of the
culprits     \citep{Kirkpatrick2024}.     Therefore,     complementary
ground-based work  on multiplicity is  essential in the Gaia  era, and
this will remain true after the final Gaia data release.

% Kirkpatrick Conclusions (9): 25%

\begin{deluxetable*}{l l cccc ccc c c}    
\tabletypesize{\scriptsize}     
\tablecaption{Spectroscopic Orbits
\label{tab:sborb}          }
\tablewidth{0pt}                                   
\tablehead{                                                                     
\colhead{HIP/} & 
\colhead{System} & 
\colhead{$P$} & 
\colhead{$T$} & 
\colhead{$e$} & 
\colhead{$\omega_{\rm A}$ } & 
\colhead{$K_1$} & 
\colhead{$K_2$} & 
\colhead{$\gamma$} & 
\colhead{rms$_{1,2}$} &
\colhead{$M_{1,2} \sin^3 i$} 
\\
\colhead{GKM} & 
& \colhead{(d)} &
\colhead{(JD -2,400,000)} & &
\colhead{(deg)} & 
\colhead{(km~s$^{-1}$)} &
\colhead{(km~s$^{-1}$)} &
\colhead{(km~s$^{-1}$)} &
\colhead{(km~s$^{-1}$)} &
\colhead{ (${\cal M}_\odot$) } 
%\colhead{} & Aa,Ab &
}
\startdata
49442 & Ab1,Ab2 & 164.55 & 60023.87 & 0.317       & 273.3    & 24.36      & 24.91    & \ldots      & 0.37 & 0.87  \\
     &    & $\pm$0.07 & $\pm$0.65  & $\pm$0.007 & $\pm$1.5 & $\pm$0.27 & $\pm$0.29 & \ldots & 0.16 & 0.85 \\
49442 & Aa,Ab & 12306 & 59624.4    & 0.285      & 335.5   & 8.09       & 4.76     & $-$12.34      & 0.65 & 0.92 \\
      &   & $\pm$1337 & $\pm$183 & $\pm$0.052 & $\pm$7.2 & $\pm$0.40 & $\pm$0.34 & $\pm$0.28  & \ldots  & 1.56 \\
55691 & Aa,Ab  & 873.14 & 60792.60 & 0.8802    &  355.83    & 13.16     & \ldots    & 3.77       & 0.024  & 0.74: \\
      &   & $\pm$0.54 & $\pm$0.05 & $\pm$0.0003 & $\pm$0.12 & $\pm$0.04 & \ldots  & $\pm$0.08   & \ldots  & 0.36 \\
61465 & Ba,Bb &  86.814 & 60049.67 & 0.233        & 130.6    & 27.62     & 35.02     & $-$13.59    & 0.058 & 1.13  \\
      &   & $\pm$0.014 & $\pm$0.16 & $\pm$0.002   & $\pm$0.7  & $\pm$0.04 & $\pm$0.06 & $\pm$0.02  & 0.093 & 0.90 \\
78662 & Ca,Cb &  0.82346 & 60556.5799 & 0     & 0    & 106.98     & 124.70     & $-$9.99    & 0.85  & 0.57  \\
      &   & $\pm$0.00001 & $\pm$0.0014 & fixed & fixed  & $\pm$1.16 & $\pm$2.09 & $\pm$0.72  & 11.33 & 0.50 \\
 GKM0 & Aa1,Aa2 & 120.74  & 60637.02    & 0.381      & 225.0      &  21.98      & 23.71     & 14.69   & 0.054 & 0.83  \\
      &       & $\pm$0.18 & $\pm$0.15 & $\pm$0.002 & $\pm$0.6  &  $\pm$0.09 & $\pm$0.09 & $\pm$0.03  & 0.075 & 0.77 \\
 GKM1 & Aa1,Aa2 & 36.589 & 57373.93 & 0.40 & 79.35 & 13.94  & \ldots & 28.92       &  0.11 &  0.71: \\
      &       & fixed & fixed 6 & fixed & fixed & $\pm$0.96 & \ldots & $\pm$0.55 & \ldots &  $>$0.21 \\
 GKM2 & Aa1,Aa2 & 11.5155 & 60779.338     & 0.030   & 91.6   & 31.51 & \ldots & 96.31       &  0.16 &  0.50: \\
      &       & $\pm$0.0002 & $\pm$0.432 & $\pm$0.003 & $\pm$13.6 & $\pm$0.26 & \ldots & $\pm$0.08 & \ldots &  $>$0.28 \\
 GKM4 & Aa1,Aa2 & 305.895 & 60761.48   & 0.235      & 76.2     & 1.69      & \ldots & 41.57       &  0.01 & 1.0: \\
      &       & fixed  & $\pm$9.43 & $\pm$0.036 & $\pm$11.7 & $\pm$0.06 & \ldots & $\pm$0.05  & \ldots & $>$0.05 \\
 GKM5 & Aa1,Aa2 & 28.5546 & 57400.8229 & 0.498   & 208.56    & 10.49 & \ldots & 8.96     &  0.37 &  0.71: \\
      &       & fixed  & fixed       & fixed   & fixed & fixed  & \ldots & $\pm$0.05  & \ldots & $>$0.11 \\
GKM6 & Aa1,Aa2 & 52.0385 & 60126.228 & 0.125   & 57.73    & 17.57 & \ldots & $-$23.04     &  0.09 &  0.98: \\
      &       & fixed  & $\pm$0.026    & fixed   & fixed & $\pm$0.06  & \ldots & $\pm$0.04  & \ldots & $>$0.37 \\
\enddata 
\end{deluxetable*}

Hierarchical systems are often  discovered by variable radial velocity
(RV) in  resolved (``visual'') binary stars.   Without follow-up work,
however, parameters  of inner  subsystems remain  unknown. A  large RV
survey  of  nearby  solar-type   hierarchies  has  been  conducted  to
determine these  parameters \citep{chiron10}. It is  complemented here
by the  detailed analysis  of four additional  hierarchies, overlooked
originally.   This project,  started before  Gaia, shows  that only  a
third of spectroscopic orbits appear in the Gaia NSS \citep[non-single
  star  catalog,][]{NSS}, and  not all  Gaia spectroscopic  orbits are
correct \citep[see also][]{Bashi2022,Gosset2025}.  The reasons are the
complexity  of  the  RV  signals from  hierarchical  systems  and  
avoidance of  stars with  resolved companions  in the  NSS.

The second  part of  this paper  reports a  test of  Gaia orbits  in a
subset of hierarchies within 100 pc observed by speckle interferometry
\citep{Tok2023}.  Their  short Gaia  periods do  not match  the longer
periods of the  speckle companions, suggesting that  these systems are
at least quadruple, considering also  the distant Gaia companions with
common  proper motion  (CPM).  However,  the RV  variability could  be
caused  by  the resolved source structure,  leading to spurious
spectroscopic   orbits  \citep{Holl2023}.    A  few   ground-based  RV
measurements can  resolve this  issue, and  seven such  candidates are
tested here.  The spectral types of these stars range from G to M, and
they are called GKM for brevity.

The data and methods, similar to  those in the previous papers of this
series,  are recalled  briefly  in  Section~\ref{sec:data}.  The  four
solar-type     hierarchies    are     discussed    individually     in
Section~\ref{sec:obj}, and  Section~\ref{sec:gkm} reports the  test of
Gaia orbits in  seven potential quadruples.  A short  discussion
in Section~\ref{sec:sum} closes the paper.

%---------------------------------------------------------
\section{Data and Methods}
\label{sec:data}

% visual
\begin{deluxetable*}{l l cccc ccc}    
\tabletypesize{\scriptsize}     
\tablecaption{Visual  Orbits
\label{tab:vborb}          }
\tablewidth{0pt}                                   
\tablehead{                                                                     
\colhead{HIP} & 
\colhead{System} & 
\colhead{$P$} & 
\colhead{$T$} & 
\colhead{$e$} & 
\colhead{$a$} & 
\colhead{$\Omega_{\rm A}$ } & 
\colhead{$\omega_{\rm A}$ } & 
\colhead{$i$ }  \\
& & \colhead{(yr)} &
\colhead{(yr)} & &
\colhead{(arcsec)} & 
\colhead{(deg)} & 
\colhead{(deg)} & 
\colhead{(deg)} 
%\colhead{} &
%\colhead{} &
}
\startdata
49442  & Aa,Ab   & 33.7  & 2022.12 &     0.285        & 0.208   & 5.2  & 335.5 & 80.3 \\
     &         & $\pm$3.7 & $\pm$0.50 & $\pm$0.052  & $\pm$0.010 & $\pm$0.5  & $\pm$7.2 & $\pm$0.6 \\   
55691 & Aa,Ab  & 2.3905     & 2025.3185  & 0.8802      & 0.1591     & 77.4      & 355.8   & 56.4     \\
     &       & $\pm$0.0015  & $\pm$0.0001 & $\pm$0.0003  & $\pm$0.0003 & $\pm$0.1  & $\pm$0.1  &  $\pm$0.6 \\ 
55691 & A,B  & 414.8       & 1918.36     & 0.672      & 5.819     & 76.6      & 16.3   & 49.8     \\
    &      & $\pm$7.8    & $\pm$0.29 & $\pm$0.006  & $\pm$0.041   & $\pm$0.6  & $\pm$1.0  &  $\pm$0.6 \\ 
%     &         & $\pm$0.  & $\pm$0.0 & $\pm$0.0  & $\pm$0.0   & $\pm$0.  & $\pm$0.  &  $\pm$0. \\  
\enddata 
\end{deluxetable*}

Observations, data reduction, and orbit calculation were described in
previous  papers of  this  series  \citep[e.g.][]{chiron8}.  To  avoid
repetition, only a brief outline  is given here.  The CHIRON fiber-fed
optical spectrometer \citep{CHIRON,Paredes2021} on the 1.5 m telescope
at Cerro  Tololo was used.   This facility  is operated by  the SMARTS
consortium,\footnote{  \url{http://www.astro.gsu.edu/~thenry/SMARTS/}}
and the observations are conducted in queue mode. Reduced spectra with
a resolution of 80,000 or 28,000 (depending on the target brightness)
are  cross-correlated   with  a  binary   mask  based  on   the  solar
spectrum. The resulting cross-correlation  function (CCF) contains one
or more dips encoding the RV, the line width, and the relative flux of
each  component.   Examples of  the CCFs  are  provided in  the  next
Section.

The  orbital  elements   and  their  errors  are   determined  by  the
least-squares fits with weights  inversely proportional to the adopted
RV    errors.    The    IDL   code    {\tt   orbit}\footnote{Codebase:
  \url{http://www.ctio.noirlab.edu/\~atokovin/orbit/}              and
  \url{https://doi.org/10.5281/zenodo.61119} } was used \citep{orbit}.
When feasible, the inner and outer orbits were fitted jointly with the
help     of     {\tt    orbit3}     \citep{TL2017}.\footnote{Codebase:
  \url{http://dx.doi.org/10.5281/zenodo.321854}} Table~\ref{tab:sborb}
lists the spectroscopic  elements. Its first two  columns identify the
system and the  component's pair. Then follow the  orbital elements in
standard notation  (period $P$, epoch of  periastron $T$, eccentricity
$e$, longitude of  periastron $\omega_A$ of the  primary component, RV
amplitudes $K_1$ and $K_2$, and  the systemic velocity $\gamma$).  The
formal errors of the elements  determined by the least-squares fit are
listed in the following line.  The  last two columns give the weighted
rms residuals to  the orbit and the projected masses  $M \sin^3 i$ for
the primary and secondary components of double-lined binaries (SB2) or
the estimated  mass of the  primary star  with colons and  the minimum
secondary mass  for single-lined orbits (SB1).   Complementary data on
the  visual orbital  elements are  provided in  Table~\ref{tab:vborb},
where some  spectroscopic elements are duplicated,  expressing periods
and periastron epochs in Julian years. The position angle of the
ascending node $\Omega_A$ and the longitude of periastron $\omega_A$
refer to the primary component. 

The individual  RVs and their  residuals to  the orbits are  listed in
Table~\ref{tab:rv}, published  in full  electronically. The  RV errors
are  assigned based  on the  residuals,  and poor  data have  inflated
errors to reduce their weight in  the orbit fit.  The errors depend on
the CCF width and contrast, signal to noise ratio in the spectrum, and
blending  with other  components.  For stars  without  orbits and  for
components with constant RV, no residuals are given.

% radial velocities
\begin{deluxetable}{r l c rrr c }    
\tabletypesize{\scriptsize}     
\tablecaption{Radial Velocities and Residuals (fragment)
\label{tab:rv}          }
\tablewidth{0pt}                                   
\tablehead{                                                                     
\colhead{HIP} & 
\colhead{System} & 
\colhead{Date} & 
\colhead{RV} & 
\colhead{$\sigma$} & 
\colhead{(O$-$C) } &
\colhead{Comp.}  \\
\colhead{GKM} & & 
\colhead{(JD -2,400,000)} &
\multicolumn{3}{c}{(km s$^{-1}$)} &
%\colhead{Instr.}
}
\startdata
 49442 & Ab1,Ab2 & 58194.6041 &    $-$35.743 &      0.350 &      0.021 & a \\ 
 49442 & Ab1,Ab2 & 58194.6041 &      9.858 &      0.350 &     $-$0.057 & b \\  
 49442 & Aa,Ab   & 58194.6041 &    $-$10.837 &      0.400 &      0.078 & c \\  
 49442 & Ab1,Ab2 & 59923.8602 &     $-$9.487 &      0.350 &      0.680 & a \\ 
 49442 & Ab1,Ab2 & 59923.8602 &    $-$26.735 &      0.350 &     $-$0.162 & b \\ 
 49442 & Aa,Ab   & 59923.8602 &     $-$3.289 &      1.000 &     $-$1.049 & c   
\enddata 
\tablenotetext{}{(This table is available in its entirety in
  machine-readable form). 
}
\end{deluxetable}

\section{Solar-Type Stars}
\label{sec:obj}

The four solar-type hierarchical systems studied here are presented in
Table~\ref{tab:objects}. The  data are collected from  Simbad and GDR3
\citep{Gaia3}, the RVs are determined  in this work.  The first column
gives the Washington Double Star \citep[WDS,][]{WDS} code based on the
J2000  coordinates.   The  HIP  and HD  identifiers,  spectral  types,
photometric and astrometric data refer  either to the individual stars
or  to the  unresolved subsystems.   Parallaxes potentially  biased by
subsystems are  marked by  colons or  relaced by  dynamical parallaxes
computed from the orbits; asterisks indicate proper motions (PMs) from
\citet{Brandt2021}.

In the  RV plots  in this  Section, green  squares denote  the primary
component, blue  triangles denote  the secondary component,  while the
full and dashed  lines plot the orbit. Typical error  bars are smaller
than  the  symbols.   Masses  of stars  are  estimated  from  absolute
magnitudes    using    standard     main-sequence    relations    from
\citet{Pecaut2013}.   Orbital  periods  of wide  pairs  are  evaluated
statistically  from  their projected  separations  \citep[see][]{MSC}.
Semimajor  axes of  spectroscopic  subsystems are  computed using  the
third Kepler's law.

\begin{deluxetable*}{c c rr   l cc rr r c }
\tabletypesize{\scriptsize}     
%\tabletypesize{\tiny}     
\tablecaption{Basic Parameters of Observed Multiple Systems
\label{tab:objects} }  
\tablewidth{0pt}                                   
\tablehead{                                                                     
\colhead{WDS} & 
\colhead{Comp.} &
\colhead{HIP} & 
\colhead{HD} & 
\colhead{Spectral} & 
\colhead{$V$} & 
\colhead{$V-K_s$} & 
\colhead{$\mu^*_\alpha$} & 
\colhead{$\mu_\delta$} & 
\colhead{RV} & 
\colhead{$\varpi$\tablenotemark{a}} \\
\colhead{(J2000)} & 
 & &   &  
\colhead{Type} & 
\colhead{(mag)} &
\colhead{(mag)} &
\multicolumn{2}{c}{ (mas yr$^{-1}$)} &
\colhead{(km s$^{-1}$)} &
\colhead{(mas)} 
%\colhead{} &
%\colhead{} &
%\colhead{} &
}
\startdata
% 49443 78662
10056$-$8405    &  A & 49442  & 88948  & F8V    & 7.00 & 1.69  & $-$107* &  0*& $-$12.3  & 13.49: DR3 \\
                &  B & \ldots & \ldots & \ldots & 8.38 & 2.05  & $-$110 & $-$3  & $-$13.8  & 13.91 DR3 \\
11247$-$6139    &  A & 55691  & 99279  & K5V    & 7.54 & 2.99  & $-$499* & 77*  &   3.8    & 86.2 dyn \\
                &  B & \ldots & \ldots & \ldots & 8.60 & 3.44  & $-$557 & 89    &   5.5    & 85.61 DR3 \\
12357$-$1201    &  A & 61466  & 109556 & G0     & 7.88 & 1.08  & $-$147 & 18    & $-$11.8  & 11.72 DR3 \\
                &  B & 61465  & \ldots & \ldots & 8.27 & 1.25  & $-$145* & 20*  & $-$13.6  & 11.49: DR3 \\
16035$-$5747    & AB &  78662 & 143474 & A7IV   & 4.63 & 0.53  & $-$120 & $-$78 & $-14.0$  & 24.67 dyn \\
                & C  & \ldots & \ldots & \ldots & 8.02 & 2.00  & $-$117 & $-$85 & $-10.0$  & 24.31 DR3 \\
\enddata
\tablenotetext{}{Proper  motions  and  parallaxes are  from  Gaia  DR3
  \citep{Gaia3}. Colons  mark  parallaxes
  biased by subsystems, asterisks mark PMs from \citet{Brandt2021}.  }
%\tablenotetext{b}{ Hipparcos parallax \citep{HIP2}.}
\end{deluxetable*}

\subsection{HIP 49442 (Quadruple)}

\begin{figure}[ht]
\plotone{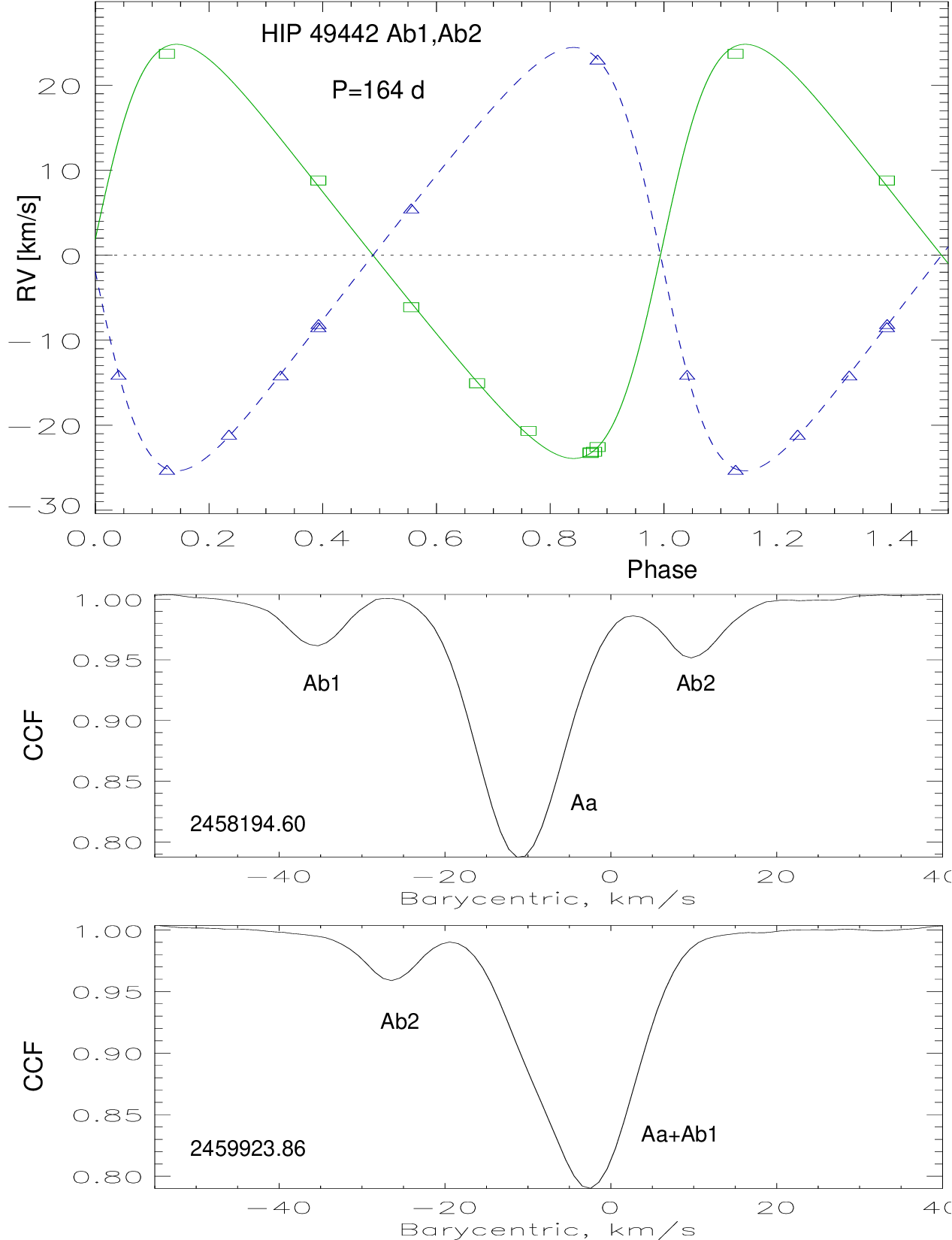}
%\epsfscale{1.1}
%\plotone{HIP49442.eps}
\caption{Spectroscopic orbit of HIP 49442 Ab1,Ab2 (top, motion in the
  orbit of Aa,Ab is subtracted) and two
  representative CCFs of star A recorded in 2018 and 2022.
\label{fig:49442} 
}
\end{figure}

\begin{figure}[ht]
\plotone{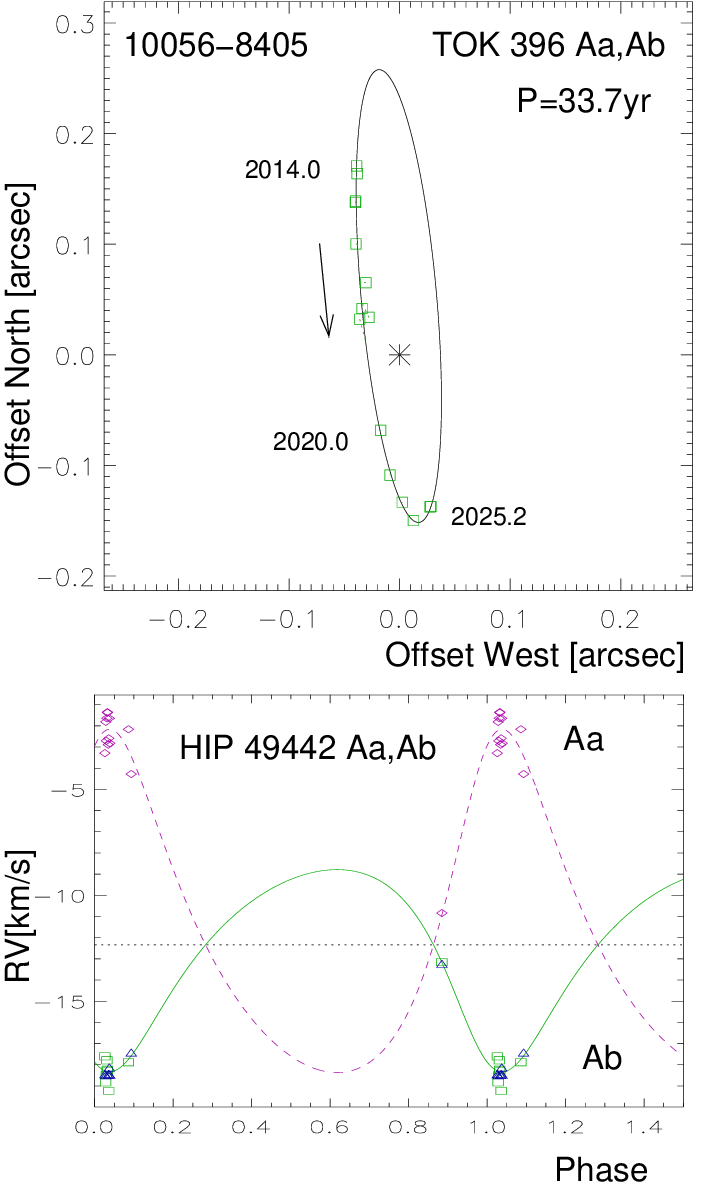}
%\plotone{HIP49442out.eps}
\caption{Orbit of HIP 49442 Aa,Ab: speckle measurements (top) and RVs
  (bottom). In the RV plot, the green line corresponds to the center
  of mass of Ab, the RVs of Ab1 and Ab2 are plotted by the squares and
  triengles with the inner orbit subtracted. The dashed magenta curve
  and diamonds depict the RVs of Aa. 
\label{fig:49442out} 
}
\end{figure}

\begin{deluxetable*}{c c r rrr rr }

\tabletypesize{\scriptsize}
%\tablenum{7}
\tablewidth{0pt}
\tablecaption{Positional Measurements and Residuals \label{tab:obs}}
\tablehead{
\colhead{HIP} & 
\colhead{System} & 
\colhead{$t$} &
\colhead{$\theta$} & 
\colhead{$\rho$} &
\colhead{$\sigma$} & 
\colhead{O$-$C$_\theta$} & 
\colhead{O$-$C$_\rho$} \\
& & 
\colhead{(yr)} & 
\colhead{(\degr)} &
\colhead{($''$)} & 
\colhead{($''$)} & 
\colhead{(\degr)} &
\colhead{($''$)} 
}
\startdata
 49442 & Aa,Ab & 2025.2064 &    191.3 &   0.1403 &   0.0020 &     -0.4 &   0.0002  \\
 49442 & Aa,Ab & 2025.2064 &    191.6 &   0.1400 &   0.0020 &     -0.1 &  -0.0001 \\
 55691 & Aa,Ab &  2024.1541 &    255.1 &   0.2987 &   0.0010 &     -0.2 &   0.0001 \\
 55691 & Aa,Ab &  2025.0892 &    272.9 &   0.1340 &   0.0010 &     -0.0 &  -0.0017 \\
 55691 & Aa,Ab &  2025.1898 &    287.8 &   0.0778 &   0.0090 &      6.4 &  -0.0077 \\
 55691 & Aa,Ab &  2025.2064 &    282.8 &   0.0756 &   0.0050 &     -1.2 &  -0.0001 \\
 \enddata
\tablenotetext{}{(This table is available in its entirety in
  machine-readable form) }
\end{deluxetable*}

This nearby  (72\,pc) star  was resolved by
J.~Herschel  in 1837  into a  3\arcsec ~visual  binary (HJ~4310)  with
$\Delta V  = 0.70$ mag. The  estimated period of A,B  is about 3\,kyr,
and  in  two centuries  the  separation  has increased  to  4\farcs35.
The astrometric acceleration of the brighter component A has been detected
by Hipparcos and confirmed  by Gaia (RUWE of 9.2 and 0.9  for A and B,
respectively);  \citet{Brandt2021} determined  a large  PM anomaly  of
$(2.7, 25.2)$  \masyr.  The  speckle-interferometric survey  of nearby
stars with accelerations  resolved in 2014 star A as  a 0\farcs18 pair
TOK~396 Aa,Ab  \citep{SOAR2014}. The  magnitude difference  between Aa
and Ab is  1.33 and 1.38 mag  in bands $I$ and  $y$, respectively. The
pair has closed down below the  resolution limit after 2018 and opened
up again in 2020, after  passing through a conjunction.  A preliminary
35-yr visual  orbit of  Aa,Ab  computed  by the
author in  2022 is  updated here using both  position measurements
and RVs.

The RV of this object was found to be variable by \citet{N04}, without
discrimination  between visual  components.   The spectrum  of star  A
taken in 2018 with CHIRON  revealed triple lines, indicating that star
Ab  is  a close  pair  Ab1,Ab2  (Figure~\ref{fig:49442}).  Thus,  this
system is a  quadruple of 3+1 hierarchy.  The object  was neglected in
the following years, and its observations with CHIRON resumed in 2022.
In the meantime, the RV of Aa  has increased from $-10.8$ to $-2$ \kms
owing to its  motion in the intermediate orbit.  Its  dip now overlaps
with the weaker dips  of Ab1 or Ab2, and for  this reason most spectra
look double-lined. The RV of star  B, measured three times, appears to
be constant at $-13.8$ \kms.  The 4\farcs35 separation between A and B
requires  accurate guiding  to prevent  mixing of  their light  in the
fiber.

Determination of  the orbit  of Ab1,Ab2  is complicated  because both
dips are equal. We  do not know which of the two  weak dips is blended
with Aa  and which is  separated.  Regular monitoring helped  to 
 associate the dips with correct components and to determine an
orbit with a period of 164 days was derived (Figure~\ref{fig:49442}).

The final iteration on  the orbits of Aa,Ab and Ab1,Ab2  was made by a
joint fit  using {\tt orbit3}.  The  orbit of Aa,Ab is  illustrated in
Figure~\ref{fig:49442out},  and  its  visual  elements  are  given  in
Table~\ref{tab:vborb}.  The position measurements  are made by speckle
interferometry  at the  4.1 m  Southern Astrophysical  Research (SOAR)
telescope. The data  before 2024 are published  in \citep{Tok2024} and
previous papers of  this series.  More recent  measurements are listed
in  Table~\ref{tab:obs}  (its  electronic version  contains  all  SOAR
measurements).    The  partial   speckle   and   RV  coverage   leaves
considerable uncertainty in  the period and other  elements.  The mass
sum of  2.7 \msun derived from  the Aa,Ab orbit and  the unbiased GDR3
parallax of star  B, 13.91\,mas, agrees with  the spectroscopic masses
of 1.0 and 1.68 \msun for Aa and Ab, respectively.

The $V$ magnitudes of Aa, Ab1, and Ab2 are 7.95, 10.08, and 10.08 mag,
respectively, based  on the speckle  photometry and assuming  that Ab1
and Ab2 are  equal.  The spectroscopic masses $M \sin^3  i$ of Ab1 and
Ab2  (Table~\ref{tab:sborb}) match the masses estimated from the absolute
magnitudes.   This implies  that  the  orbit of  Ab1,Ab2  has a  large
inclination  and suggests  its possible  alignment with  the orbit  of
Aa,Ab.  However,  the estimated mass  of Aa, 1.26 \msun,  which agrees
with its spectral type  F8V, is larger than 1 \msun derived from the
combined orbit  of Aa,Ab.  Further observations  will eventually yield
more accurate measurement of all masses.

The angular  semimajor axis  of Ab1,Ab2 is  9.8\,mas, so,  despite the
large inclination,  the innermost pair  can be marginally  resolved by
speckle  interferometry  at 8  m  telescopes.   Such a resolution  would
determine the mutual inclination between  Aa,Ab and Ab1,Ab2.  Owing to
the near-equality of Ab1 and Ab2, the wobble in the outer orbit is too
small to be detectable. The residuals of the speckle measures of Aa,Ab
to the orbit  are about 1\,mas. The relatively large  residuals in the
RVs of Aa are caused by frequent blending of its CCF dip with the dips
of Ab1 or Ab2.

The lithium 6707\AA  ~line is clearly detectable in the  spectra of Aa
and B, indicating that  this system is juvenile. The dips  of Aa and B
are slightly  widened by  rotation ($V  \sin i$ or  6.8 and  4.0 \kms,
respectively),  while the  lines of  Ab1 and  Ab2 do  not exhibit  any
measurable rotational broadening.

% No X-rays. Variability? 
% Lithium in Aa and B
%U,V,W:   -34.36    4.46  -16.99
% speckle/doc/SAM23/HIP49442/memo.pdf
% msc/triple/orbits2/10056-8405.in3
% Simbad: checked references, nothing relevant
% Not found in the paper by Menzez et al. (2025) on Zorro.

\subsection{HIP 55691 (Triple)}

\begin{figure}[ht]
\plotone{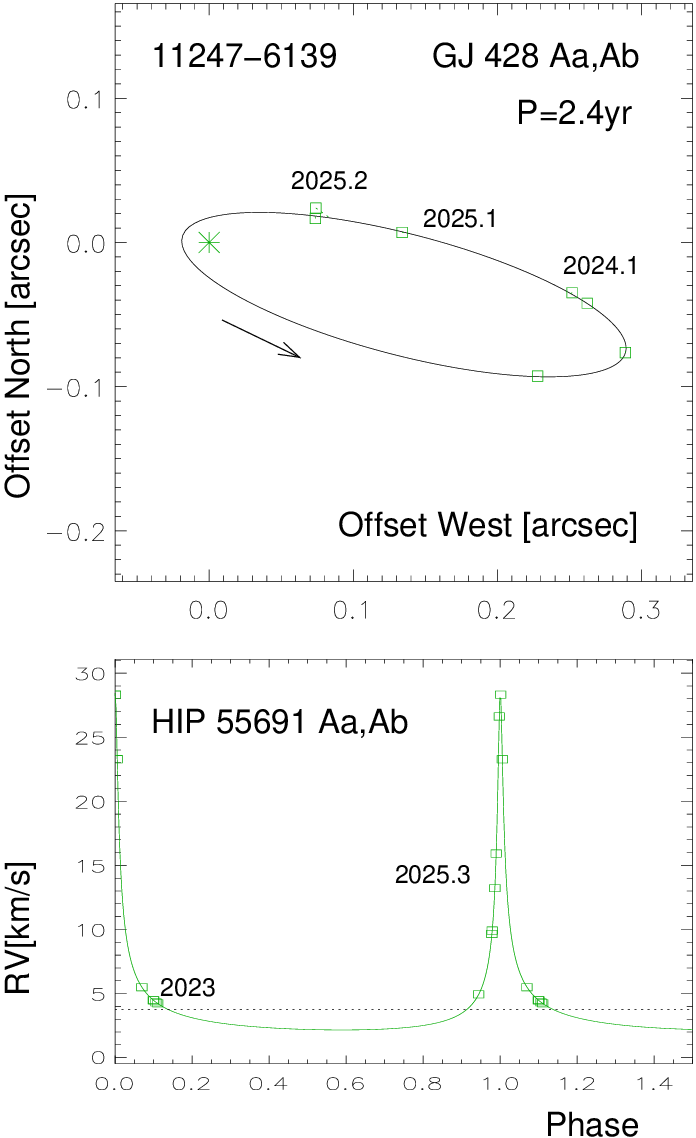}
%\epsfscale{1.1}
%\plotone{HIP55691in.eps}
\caption{Visual orbit (top) and the RV curve (bottom) of HIP 55691 Aa,Ab.
\label{fig:55691} 
}
\end{figure}

This triple system is a close  neighbour located at 11.7\,pc. The main
star  A  is known  as  HIP~55691 or
GJ~428.  The visual companion B has been noted by J.~Hershel in 1834 at
4\arcsec. Its  motion, monitored for almost  two centuries, constrains
the outer orbit  quite well. The pair A,B
passed through the  periastron in 1918 and slowly opens  up, now being
separated by 7\farcs6.

A strong suspiction that star A contains a subsystem was raised by the
acceleration, namely  the PM  anomaly detected via  comparison between
Gaia DR2 and Hipparcos by \citet{Brandt2018} and \citet{Kervella2019}.
The lack of 5-parameter astrometry in GDR3 indicates that the star was
probably resolved.  Its  first observation at SOAR in  2021 revealed a
faint ($\Delta  I_{\rm Aa,Ab}  = 3.0$ mag)  companion Ab  at 0\farcs25
separation    \citep{Tokovinin2022}.    Subsequent    monitoring   was
inconclusive until  2025, when the  pair became closer and  started to
move faster.   The speckle  data suggested an  eccentric orbit  with a
period  of 2.4  yr.   Such an  orbit would  produce  a substantial  RV
variation at periastron, predicted  for 2025 April.  Spectroscopy with
CHIRON in  2025 confirmed this  prediction and enabled  calculation of
the    combined   spectro-interferometric    orbit   illustrated    in
Figure~\ref{fig:55691}. The  eccentricity of  $e_{\rm Aa,Ab}  = 0.8802
\pm 0.0003$ is high, albeit not  extreme.  Five CHIRON spectra of this
star taken  in 2023 by  T.~Johns in a survey  of nearby K  dwarfs were
also used (the RVs were measured by cross-correlation).

The mass  of Aa deduced  from its  absolute magnitude, 0.74  \msun, is
normal  for  a  K5V  dwarf.   The RV  amplitude  and  the  inclination
correspond to  the mass of  Ab of  0.36 \msun, or  a mass sum  of 1.10
\msun; the mass  sum computed from the orbit and  the parallax is 1.12
\msun.   The  amplitude of  the  photocenter  wobble should  be  about
43\,mas; its ratio  to the semimajor axis is the  wobble factor $f^* =
q/(1+q) -  r/(1+r) = 0.27$, where  $q_{\rm Aa,Ab} = 0.49$  is the mass
ratio and $r_{\rm Aa,Ab} = 0.06$  is the flux ratio corresponding to a
magnitude difference of 3 mag.  No astrometric or spectroscopic orbits
are provided  in the GDR3,  possibly because of  insufficient sampling
near  the  periastron.   However,  the  wobble  caused  by  the  inner
subsystem seriously biases the astrometry  of star A in both Hipparcos
and Gaia missions.   The best (but not perfect) estimate  of the PM of
A, $(-499.40,  77.46)$ \masyr, is obtained  by \citet{Brandt2018} from
the Hipparcos and GDR2 positions.

\begin{figure}[ht]
\plotone{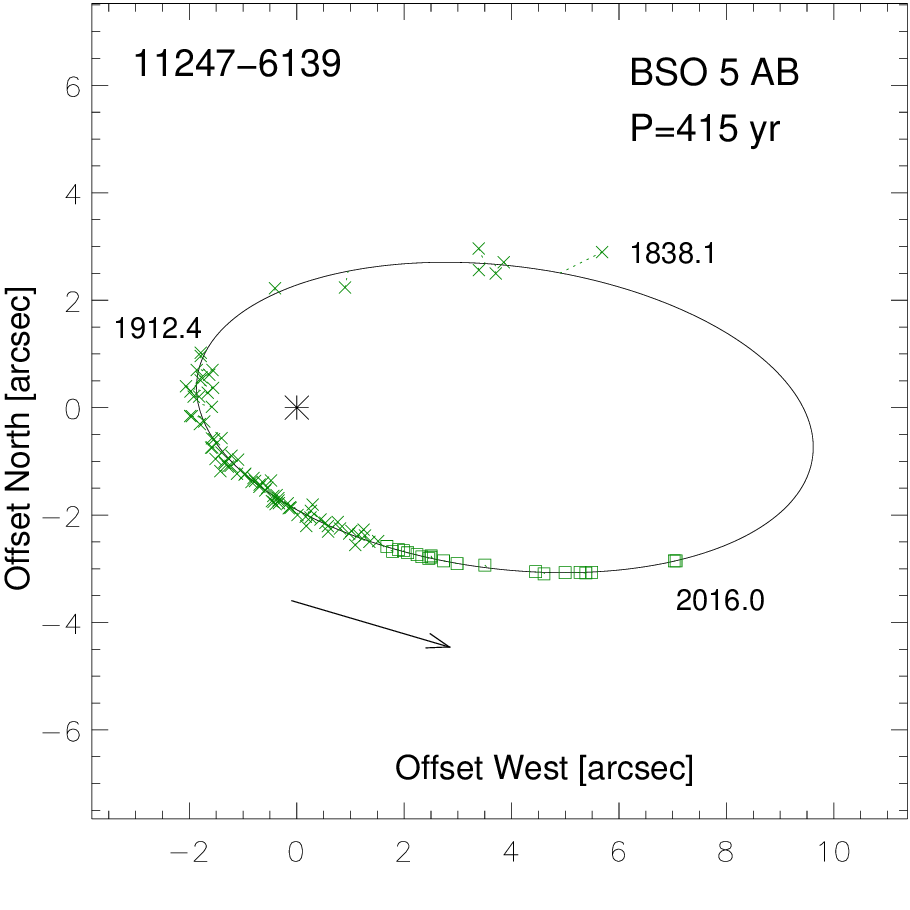}
%\plotone{HIP55691out.eps}
\caption{Visual  orbit of  HIP 55691  A,B (WDS  12247$-$6139). Crosses
  mark  micrometer  measurements,  and squares  stand  for  more  accurate
  photographic measurements and space astrometry.
\label{fig:55691out} 
}
\end{figure}

\begin{figure}
\epsscale{1.0}
\plotone{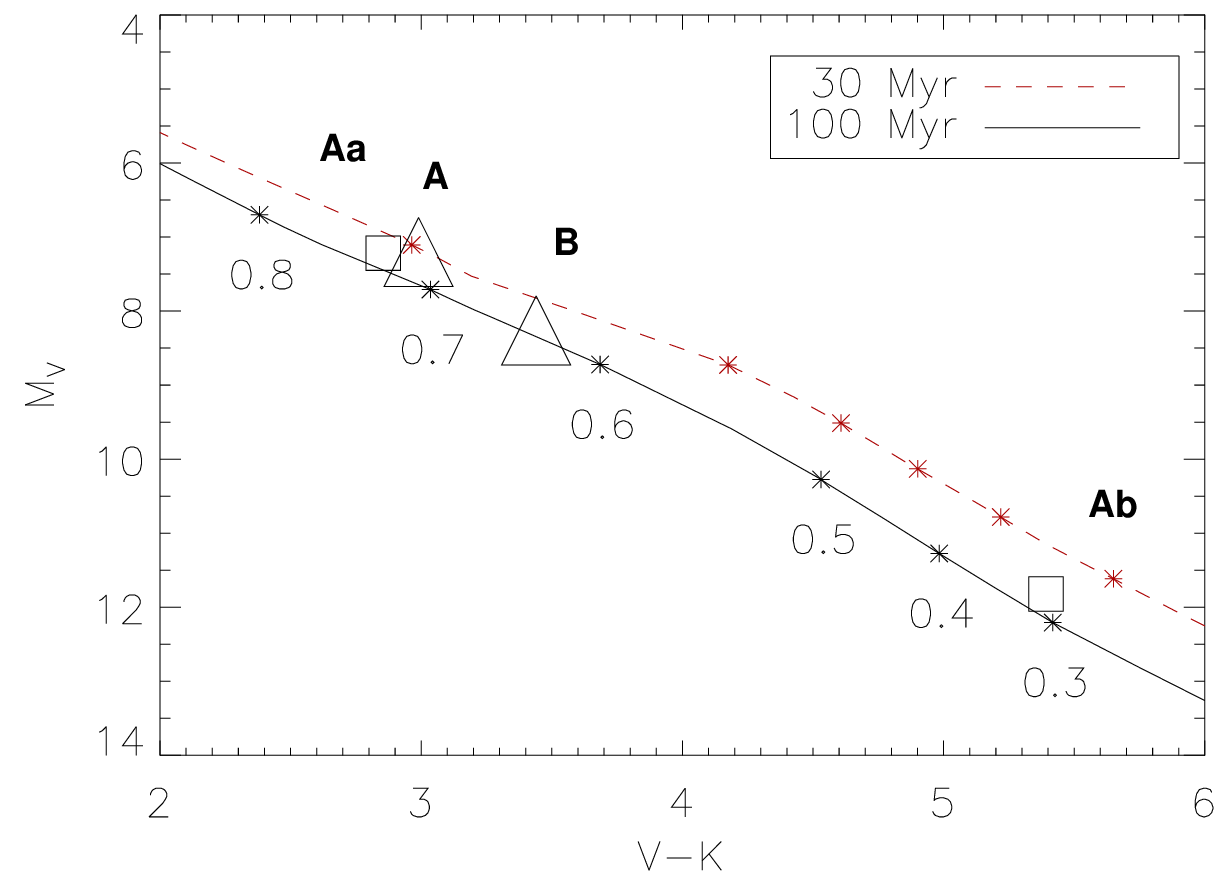}
%\plotone{CDM.eps}
\caption{Color-magnitude diagram  of HIP 55691. The  PARSEC isochrones
  for 100  and 30 Myr are  plotted, with asterisks marking  masses
  from 0.3 to 0.8 \msun.  The positions of stars A and B are marked by
  large triangles, while smaller  squares indicate tentative locations
  of Aa and Ab.
\label{fig:CMD55691}
}
\end{figure}

The orbit of the outer pair  A,B has been computed and updated several
times in the past.  I refined its elements (see Table~\ref{tab:vborb})
by assigning  appropriate errors  (hence weights)  of 0\farcs2  to the
micrometer  measurements, 0\farcs05  to the  photographic measurements
after  1950  (the micrometer  data  after  1950,  as well  as  obvious
outliers, are ignored), 10\,mas to  the Hipparcos position, and 2\,mas
to the  relative position in  GDR3.  This relatively  well-constrained
orbit  plotted in  Figure~\ref{fig:55691out} is  similar to  the orbit
computed  by \citet{Izmailov2019}.   The  weighted  rms residuals  are
10\,mas.  Using the accurate GDR3  parallax of star B, 85.61$\pm$0.021
mas, the  orbit leads  to the  mass sum  of 1.83  \msun, close  to our
estimate of 1.74 =  1.10 + 0.64 \msun.  The motion of  B relative to A
in 2015.5 was  $(-54.53 +12.47)$ \masyr according to  this orbit.  The
difference between  PMs of B and A, $(-57.4,  +11.1)$  \masyr,  is  in
reasonable agreement.  Based  on the estimated masses of A  and B, the
center of mass of the system has a PM of $(-522, +81)$ \masyr.

The systemic RV  of A, 3.77$\pm$0.07 \kms, differs from  the three RVs
of B measured with CHIRON (mean 5.50$\pm$0.02 \kms) owing to motion in
the outer orbit.   Fixing the RV amplitude $K_{\rm A}  = 1.72$ \kms to
its estimated value, I obtain $K_{\rm B} = 3.45$ \kms and the systemic
velocity of  4.32 \kms. The ascending  nodes of both orbits  are known
without  ambiguity,  and  their  mutual  inclination  is  $6\fdg5  \pm
0\fdg8$.  Similar orientation of inner  and outer orbits is notable by
comparing Figures~\ref{fig:55691}  and \ref{fig:55691out}.   The lines
of apsides of both orbits also point in the same direction. Star B has
good astrometry in GDR3 (a RUWE of  1.1) and a constant RV, so it does
not host inner subsystems, as far as we can tell.

After computing the  inner and outer orbits separately,  I fitted them
jointly using {\tt orbit3}, and  the elements given here correspond to
this final iteration. The tiny wobble in the motion of A,B  does not look
convincing  on  the  orbit  plot, but the fitted coeficient 
 $f^* = 0.19 \pm 0.03$ is formally significant and comparable to its
estimate. 

This star has been detected  in X-rays and, accordingly, was suspected
to be young  by \citet{Torres2006}.  However, the Li  6707\AA ~line is
not seen in  the CHIRON spectra of Aa  and B, and the width  of the CCF
dips corresponds  to a moderate projected  rotation $V \sin i$  of 4.3
and 5.6 \kms for A and B, respectively.  The spatial velocity $(U,V,W)
= (-24.8,  -15.2, -5.3)$ \kms  computed using the estimated  motion of
the center of mass of AB  suggests potential membership in the IC~2391
(Argus) association with an age of $\sim$50\,Myr \citep{Nakajima2012}.
In   Figure~\ref{fig:CMD55691}   the   stars   are   placed   on   the
color-magnitude  diagram and  compared  to two  PARSEC isochrones  for
solar metallicity  \citep{Bressan2012}.  The  magnitudes of Aa  and Ab
are assigned  tentatively by assuming  masses of 0.74 and  0.36 \msun,
computing the magnitude differences in the  $V$ and $K$ bands from the
100  Myr isochrone,  and  splitting the  total  flux accordingly.   As
$\Delta V_{\rm  Aa,Ab} =  4.7$ mag is  large, the  absolute magnitudes
$M_V$ of Aa and A=Aa+Ab are almost identical, but Aa is slightly bluer
than A.  The 100 Myr isochrone predicts a magnitude difference of  3.3 mag in
$I$, close  to the measured value  (3.0 mag).  On the  other hand, the
masses  and the  30 Myr  isochrone correspond  to a  smaller magnitude
difference  of $\Delta  I_{\rm Aa,Ab}  = 2.1$  mag, contradicting  the
observations.  Interestingly, the  mass of Ab measured  from the inner
orbit and the magnitude difference,  taken together, help us constrain
the age of this triple system: about 100 Myr or slightly less.  A more
refined analysis can be made in the future when relative photometry of
Aa,Ab in various bands becomes available.

\subsection{HIP 61466+61465 (Quadruple)}

\begin{figure}[ht]
\plotone{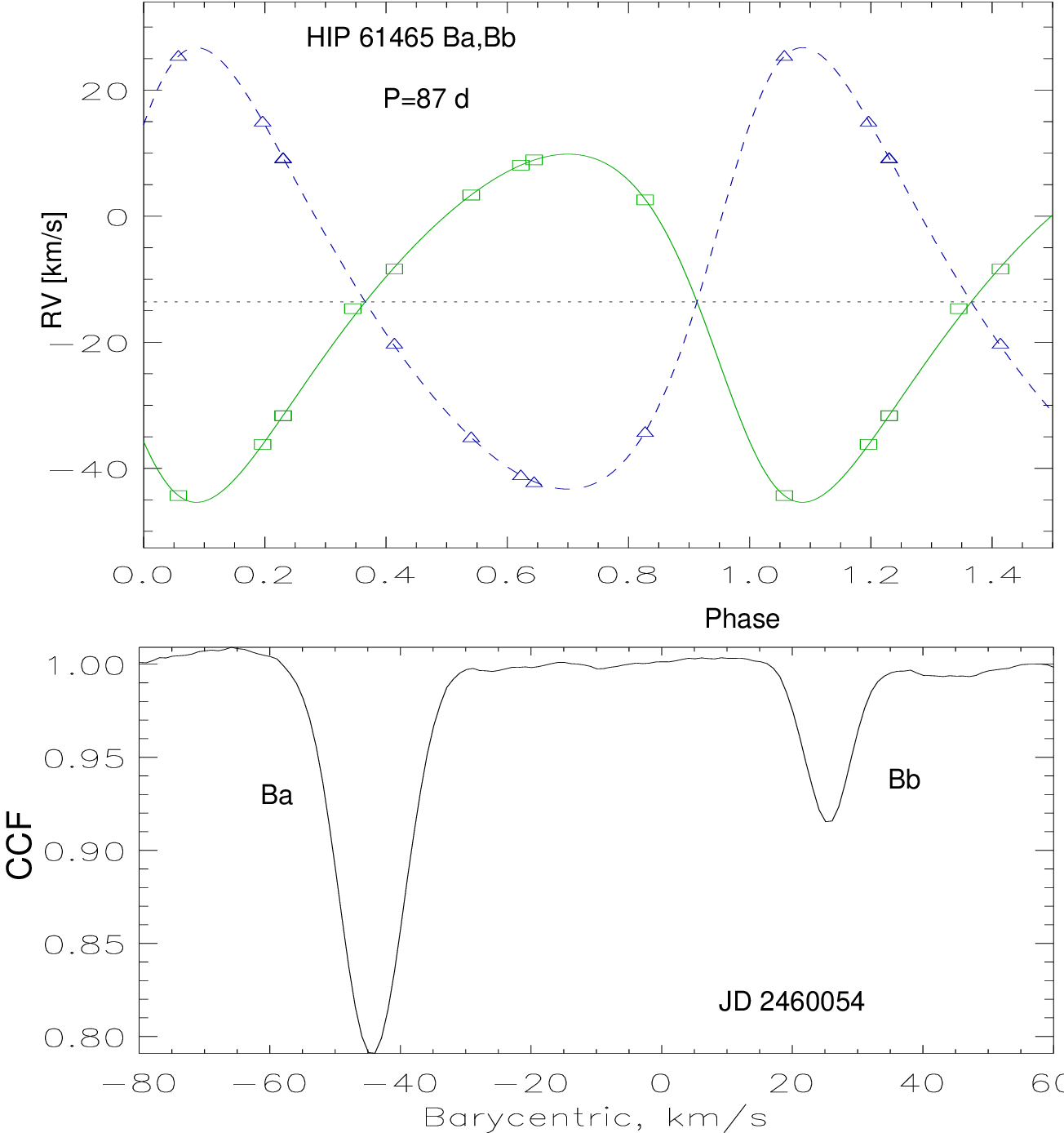}
%\plotone{HIP61466.eps}
\caption{Spectroscopic orbit of HIP 61465 Ba,Bb (top) and the CCF
  recorded on JD 2,460,054 (bottom).
\label{fig:61466} 
}
\end{figure}

This hierarchical system located at 85\,pc is composed of bright stars
A (HIP  41466) and B  (HIP 41465)  at 27\farcs5 separation.  The wide
pair STF~1659  has moved  very little since  its first  measurement by
W.~Struwe in  1828. The WDS lists  fainter components C to  F, but all
these  stars are  optical,  as evidenced  by  their unstable  relative
positions, mismatching PMs, and parallaxes.

\citet{Brandt2021}  detected a  statistically  significant PM  anomaly
(i.e. a long-term  acceleration) of 2  \masyr in both  stars, suggesting
existence of  inner subsystems.  However,  it is known  that Hipparcos
astrometry  of  visual  binaries  with separations  on  the  order  of
20\arcsec ~often has large errors  owing to its measurement procedure,
so the  veracity of the  anomaly derived from the  Hipparcos positions
remains questionable;  GDR3 has  a RUWE of  1.0 and 3.1  for A  and B,
respectively.

\citet{Desidera2006}  measured  the  RVs  of  stars  A  and  B  on  JD
2\,447\,720.3  at   $-14.77$  and  $+4.78$  \kms,   respectively,  and
suggested  that both  can be  variable; they  have not  noticed double
lines in the spectrum of B.  The RVs of A measured with CHIRON in 2018
and 2023 are $-11.83$ and $-10.77$ \kms. The slow positive RV trend of
star  A  is caused  by  a  subsystem with  a  long  but still  unknown
period. This pair,  presumably responsible for the 
acceleration, could possibly be resolved by speckle interferometry or
adaptive optics.

Double lines  in the  CHIRON spectrum  of star B  were noted  in 2018.
However,  the orbit  has not  been  pursued at  the time.   Additional
spectra taken in 2023 confirm the double-lined nature and establish the
orbit  of  Ba,Bb  with  $P=86.8$  days  (Figure~\ref{fig:61466}).  The
photocenter motion is a likely cause  of the astrometric noise in
GDR3.

Comparison between the masses of Ba and Bb estimated from their absolute
magnitudes (1.20 and 0.91 \msun) and the spectroscopic masses $M
\sin^3 i$ of 1.16 and 0.97 \msun indicates that the orbit of Ba,Bb has
a large inclination. The Li line was noted in the spectrum of the
brighter component Ba, but it is absent in the spectrum of A.

\subsection{HIP 78662 (Quadruple)}

\begin{figure}[ht]
\plotone{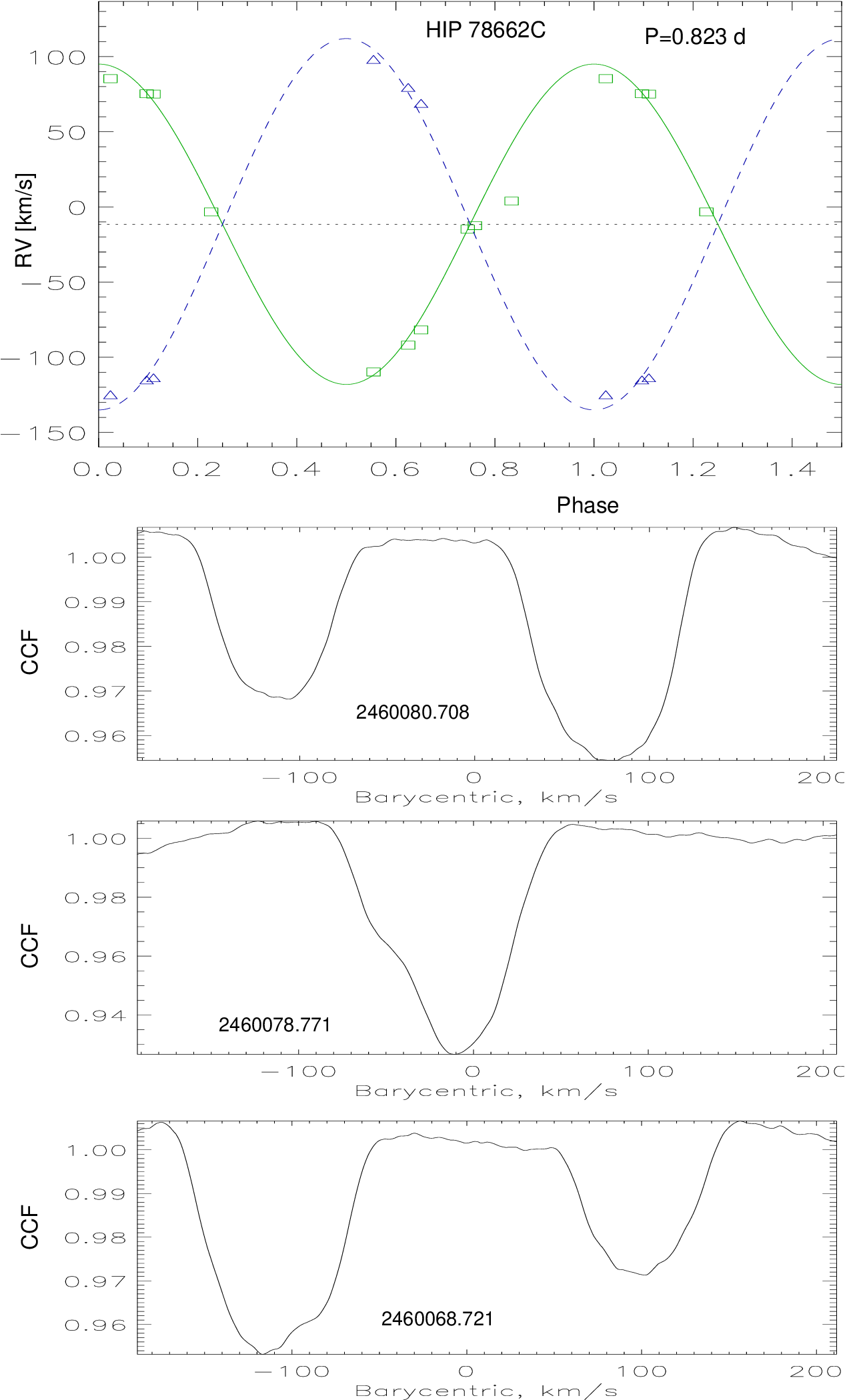}
%\plotone{HIP78662C.eps}
\caption{Spectroscopic orbit of HIP 78662 Ca,Cb (top) and three representative CCFs (bottom).
\label{fig:78662} 
}
\end{figure}

\begin{figure}
\plotone{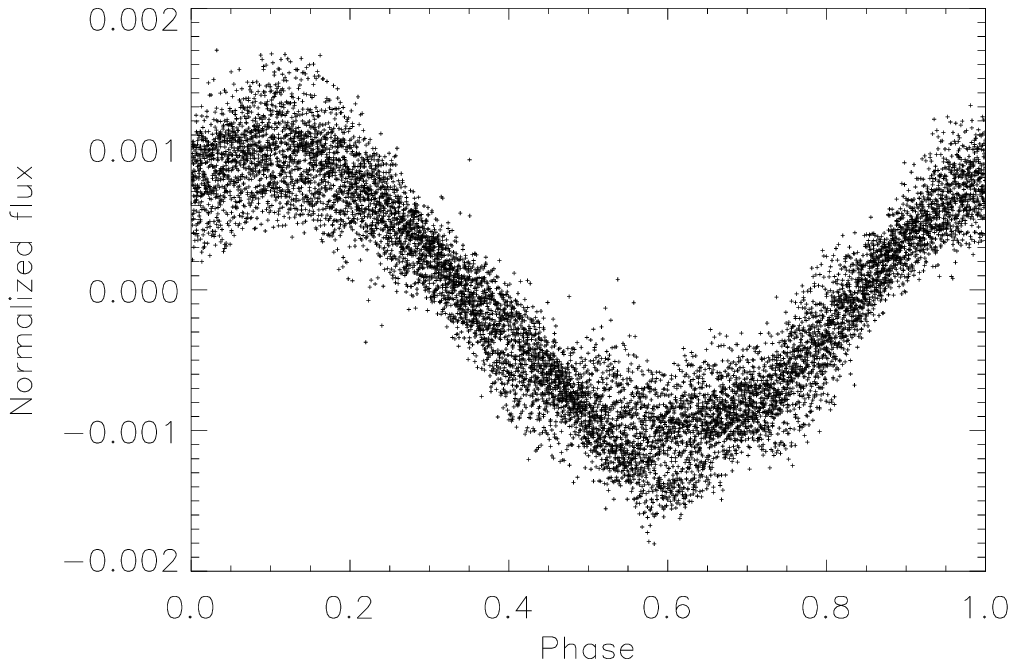}
%\plotone{HIP78662-lightcurve.ps}
\caption{Light curve of HIP~78662C from the TESS Sector 39 folded with 
 a  period of  0.81947 days and an arbitrary initial phase.
\label{fig:LC}
}
\end{figure}

This is  a bright  quadruple system.  The  main star,  $\iota$~Nor (HR
5961), has  a visual  companion C  at 11\arcsec,  first noted  by John
Hershel  in 1835  (HJ~4825).   Although star  C  is relatively  bright
($V=8.02$  mag),  it has  not  deserved  individual identifies  except
2MASS, apparently being lost in the glare of its brighter sister.  The
latter has been  resolved in 1897 as a 0\farcs8  pair SEE~258, and its
visual  orbit with  $P=26.84$ yr  is very  well known.   The dynamical
parallax  of 24.68\,mas  derived  from this  orbit  and the  estimated
masses matches  well the  GDR3 parallax  of C, 24.31\,mas.   As AB  is a
binary, Gaia did  not determine its astrometric  parameters, while the
Hipparcos parallax of 25.4 mas is less accurate.

Star C was observed with CHIRON in 2018. The spectrum revealed very
broad and low-contrast double lines. Considering  the likely short
period, the low expected accuracy of the RVs, and other targets of higher
priority, further study was postponed till 2023. New spectra confirmed
the double-lined nature of this star and its rapidly variable
RVs. Three representative CCFs in the lower panels of
Figure~\ref{fig:78662} illustrate the dips of unequal amplitudes that
swap within a few days. When the dips partially overlap, the CCFs are
broad and of irregular shape, so attempts to fit overlapping dips
failed, except when the overlap is nearly perfect and the single dip
is not too wide, as shown.  The orbital period of 0.82 days was
confirmed by taking two spectra  within a couple of hours on a single
night.

A short  orbital period suggests  possibility of eclipses.   The light
curves (LCs)  of HIP  78662C (TIC 426452677)  were retrieved  from the
TESS \citep{TESS}  archive for  sectors 1 (2018)  and 39  (2021), both
with a 2 min. cadence.  Both data sets show a low-amplitude modulation
with  dominant periods  of 0.41  and 0.82  days in  sectors 1  and 39,
respectively.  The latter has a larger  amplitude, and the LC is shown
in Figure~\ref{fig:LC}.  The LCs refer  to the combined light of three
stars where  the flux is  dominated by  the brighter components  A and
B. Although early-type (A7IV) stars A and B might have rapid rotation,
they are  unlikely to have  spots.  The light modulation  is therefore
attributable to the chromospherically active  stars Ca and Cb, tidally
synchronized with the orbit. The  modulation periods are very close to
the orbital period and its second harmonic. The irregular shape of the CCF
dips is likely caused by the starspots.

The  TESS LC  clearly shows  the lack  of eclipses.   The same  conclusion
emerges from the amplitudes of the RV variation that correspond to the
projected masses $M \sin^3 i$ of  0.37 and 0.33 \msun. Comparison with
masses  estimated  from   $M_V$,  around  0.9  \msun,   leads  to  the
inclination of $45^\circ$. Summarizing, this system is a 2+2 quadruple
with very unequal inner periods of 26.8 yr and 0.82 d. If the twin pair Ca,Cb
merges, the mass of the  resulting blue struggler, 1.75 \msun, will be
similar to the masses of stars A and B, 1.75 and 1.54 \msun
respectively.

%-------------------------------------------
\section{Testing Hierarchies Within 100 pc}
\label{sec:gkm}

\begin{deluxetable*}{l l l l  c c   c c c  l }
\tabletypesize{\scriptsize}     
%\tabletypesize{\tiny}     
\tablecaption{Stars with NSS Orbits and Speckle Companions
\label{tab:gkm} }  
\tablewidth{0pt}                                   
\tablehead{     
\colhead{GKM} & 
\colhead{WDS} &  
\colhead{DR3} &  
\colhead{Sol.} & 
\colhead{$P_{\rm NSS}$}  & 
\colhead{$K_{\rm NSS}$}  & 
\colhead{$\rho$} & 
\colhead{$\Delta I$} & 
\colhead{$P^*$} & 
\colhead{Comment} \\
  &  & & &
\colhead{(d)} &
\colhead{(\kms)} &
\colhead{($''$)} &
\colhead{(mag)} &
\colhead{(yr)} &
}
\startdata
0 &04097$-$5256 & 4780342550850239872 & AORB    & 121.9  & \ldots & 0.176  &  3.4 & 38 & SB2  \\ %0.7mas P*=45y 
1 &07543$+$0232 & 3088161818892312704 & SB1     & 36.6   & 15.2   & 0.364  &  2.3 & 155 & Confirmed \\ % P*=154yr
2 &09336$-$2752 & 5633969259436187648 & SB2     & 37.2   & 37.6   & 0.139  &  2.9 & 13 & $P=11.5$d \\ %P*=13y no contamin.
3 &15397$-$4956 & 5985420622174761600 & SB1     & 44.8   & 10.0   & 0.046  &  0.8 & 5  & RV const. \\ % K1=10.0 P*=5.3yr
4 &20147$-$7252 & 6373829259376877440 & SB1     & 305.9  & 2.2    & 0.253  &  3.5 & 75 & Confirmed \\ %K1=2.2 P*=75yr
5 &21460$-$5233 & 6462355265559115392 & SB1     & 28.6   & 10.5   & 0.088  &  2.5 & 16 & Confirmed \\ % K1=10.5  P*=15.8yr 
6 &12022$-$4844 & 6130627011126533760 & SB1     & 52.0   & 16.7   & 0.095  &  3.1 & 13 & Confirmed \\
\enddata
\end{deluxetable*}

A large speckle survey of candidate hierarchical systems within 100 pc
derived from the Gaia GCNS  \citep{GCNS} enabled confirmation of a few
hundred triples where the wide outer pair consists of two
Gaia sources, and the inner  pair, resolved by speckle interferometry,
is  revealed by the Gaia  multiplicity  indicators \citep{Tok2023}.   In
several cases, the NSS \citep{NSS}  contains orbits of resolved inner
pairs.  Twenty such systems are listed in Table~3 of \citet{Tok2023}.

Some of the NSS binaries coincide with the close speckle pairs because
the orbital periods  inferred  from the separation, $P^*$, match the NSS
periods. These  cases are  confirmed by subsequent  speckle monitoring,
allowing    calculation    of   several   visual    (resolved)    orbits
\citep{Tok2024}.  However,  in the  remaining  cases  the NSS  periods
$P_{\rm  NSS}$ are  much shorter  than $P^*$.  If the  NSS orbits  are
verified, these would  be quadruple systems of 3+1  hierarchy (a close
NSS  pair,  an intermediate  speckle  pair,  and the  outer  companion
resolved by Gaia).

 However, the NSS spectroscopic orbits of close pairs require confirmation.
Gaia  measured the  RVs by  a  slitless spectrometer  with a  spectral
resolution of 11,500. The RV depends  on the source position along the
scan determined  by the  astrometric solution;  an offset  of 1\arcsec
~corresponds  to  an RV  offset  of  146  \kms.  Presence of  a  close
companion  can  bias  the  RVs  in  several  ways,  leading  to  false
spectroscopic orbits, as described in section~3.3 of \citet{Holl2023}.
A  companion separated  by $\rho$  from  0\farcs1 to  0\farcs9 can  be
resolved in  scans along the  binary direction,  and in such  case the
Gaia astrometry refers  to the main star (which also  dominates in the
spectrum). However,  in other scans  that are nearly  perpendicular to
the binary, the components are blended  in the common Gaia window, and
the astrometry  refers to  the photocenter.   Mixture of  resolved and
unresolved scans degrades  the quality of Gaia  astrometry, causing an
increased RUWE.   The RVs are also  biased.  If the flux  ratio in the
Gaia radial velocity spectrometer band  (at 870\,nm) is $r$, the shift
between the primary  star position and the photocenter is  $\rho r /(1
+r)$, and the corresponding RV bias  can reach a few \kms.  An example
of spurious  NSS SB1  orbit with  an amplitude of  $\sim$9 \kms  for a
0\farcs3  binary  is  given  in Figure~14  of  \citet{Holl2023}.   For
unresolved  point-like  sources  such  as  astrometric  binaries,  the
photocenter  motion  (not accounted  for  in  the Gaia  spectroscsopic
orbits) also  produces RV errors  and biases the elements,  albeit not
dramatically.

The  Gaia   scan  direction   changes  with   a  period   of  $\sim$63
days. Spurious  signals caused  by the  source structure  might appear
periodic,  with a  range of  possible periods;  the 31.5  day spurious
period is  one of the strongest  \citep{Holl2023}.  Speckle companions
(except the closest) move little during the 3 yr time span of GDR3, so
the companion-induced RV  errors should be related mostly  to the scan
direction.

If some NSS orbits of stars with speckle companions are spurious, they
are not 3+1 quadruples, just triples. To test the orbits, RVs of seven
candidate stars  were monitored  with CHIRON.   These stars  are given
here short  names from GKM0  to GKM5. The  last object, HIP  58691 (HD
104532), is  relatively bright  ($V=9.26$ mag); it  is denoted  here as
GKM6.  The WDS-like  codes and other data of these  stars are reported
in Table~\ref{tab:gkm}.  It  contains the Gaia NSS  solution type, its
period $P_{\rm  NSS}$, and  the RV  semiamplitude $K_{\rm  NSS}$.  The
following columns give the separation $\rho$, the magnitude difference
$\Delta I$, and the estimated periods $P^*$ of the speckle companions.
The  separations and  magnitude  differences  correspond to  potential
spurious RV  amplitudes $ 146  \; \rho  r/(1+r)$ of 1--2  \kms, except
GKM0 where it reaches 5.7 \kms. The visual magnitudes of the first six
stars range from 9.4 to 12.6, and they were observed in the fiber mode
with  a spectral  resolution of  28,000, while  HIP~58691 (GKM6)  was
observed in the slicer mode with a resolution of 80,000.

\begin{figure*}
\epsscale{1.1}
\plotone{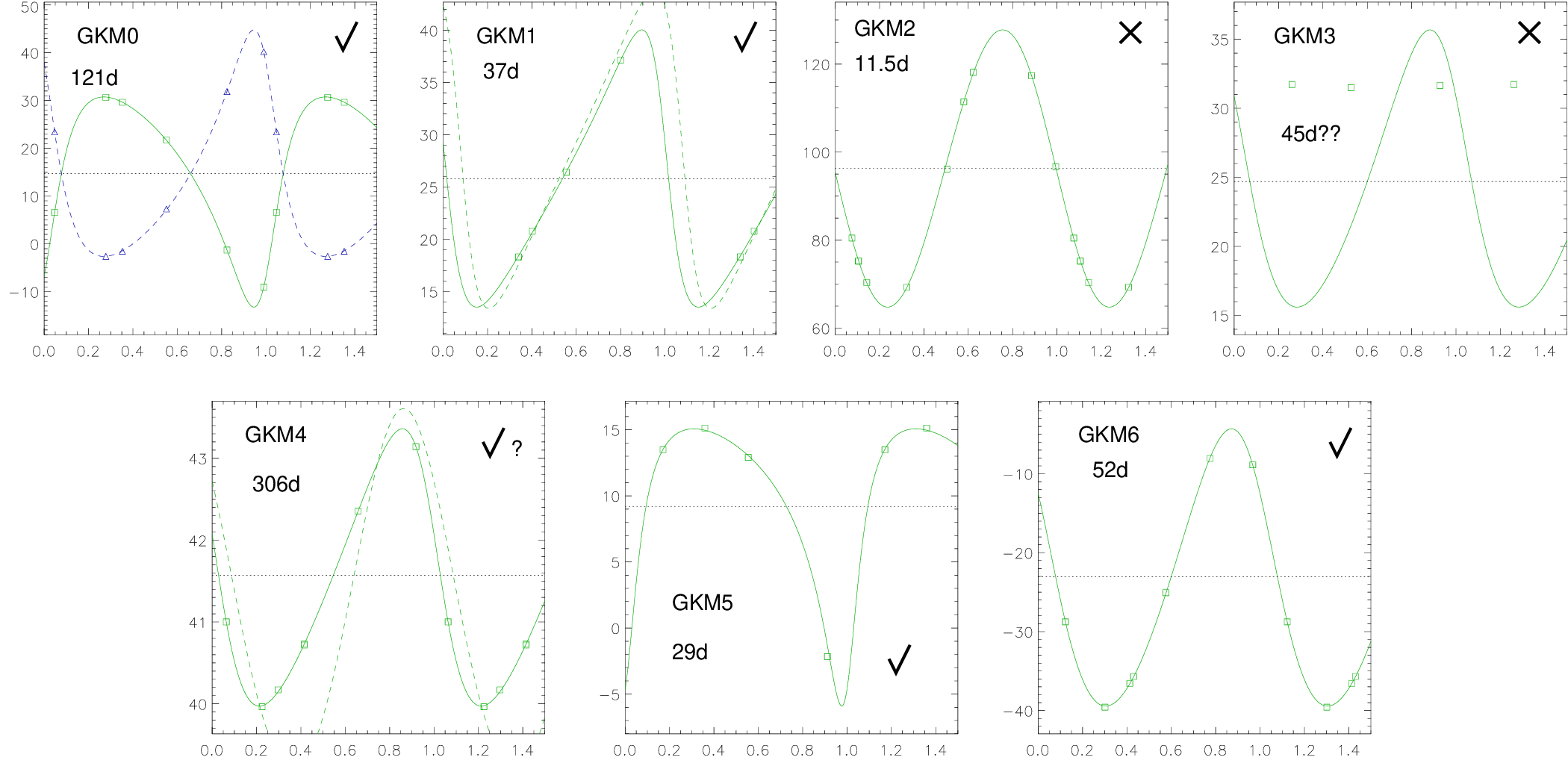}
%\plotone{GKMorb.eps}
\caption{RV  curves  of seven  GKM  stars  observed with  CHIRON  (see
  text). Confirmed Gaia  periods are marked by  ticks, unconfirmed ---
  by crosses. The RVs of  primary and secondary components are plotted
  by squares  and triangles, respectively,  the fitted orbits  by full
  lines, and some GDR3 orbits by dashed lines.
\label{fig:GKMorb}
}
\end{figure*}

\begin{figure*}
\epsscale{1.0}
\plotone{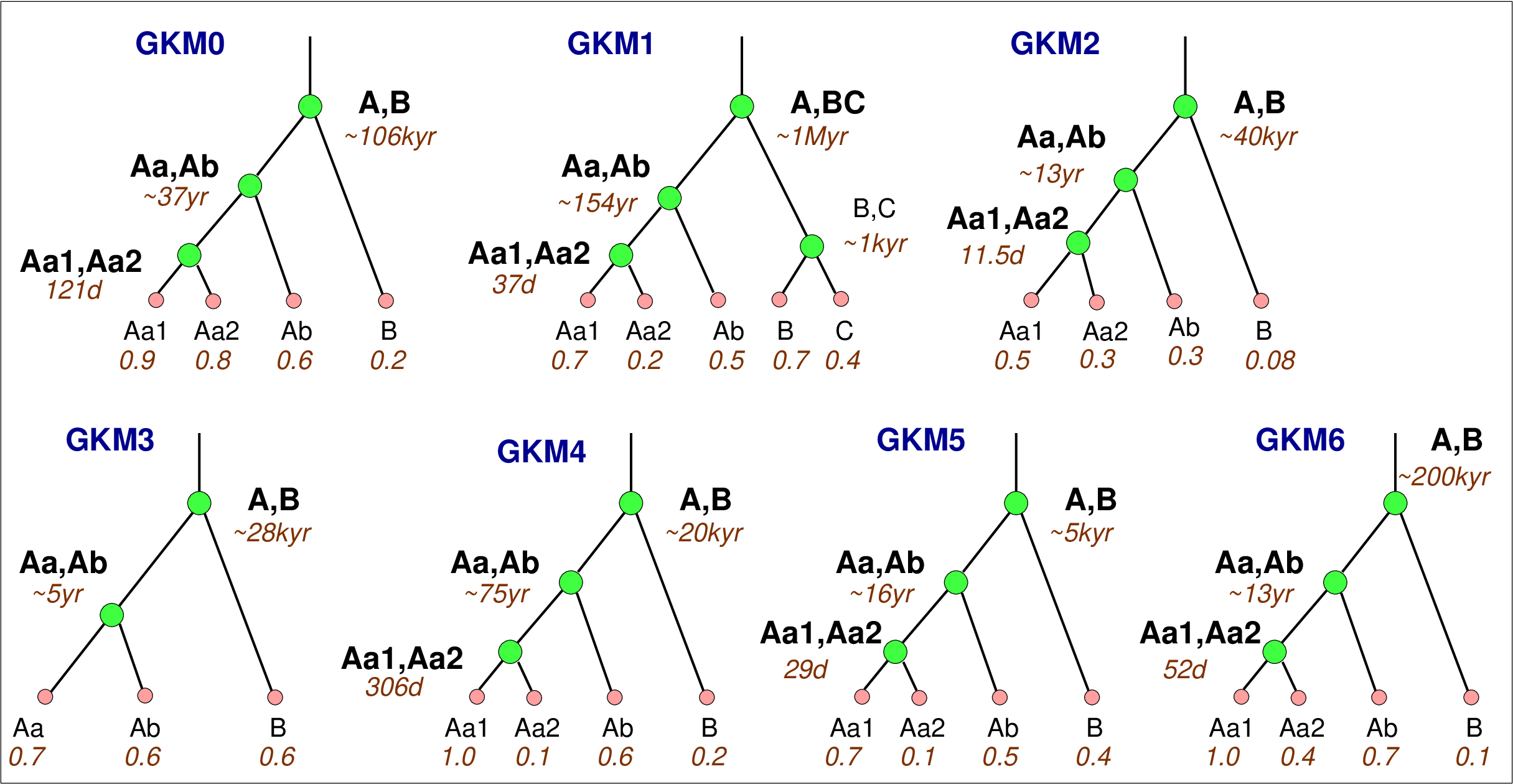}
%\plotone{mobile.eps}
\caption{Architecture  of  seven  GKM  systems.   Green  circles  show
  subsystems (their known or estimated  periods are given), small pink
  circles  --- individual  stars; their  approximate masses  are shown
  below.
\label{fig:mobile}
}
\end{figure*}

The original intent was  to measure only a few RVs of  each star for a
quick verification of  the GDR3 orbits. Some  stars, however, required
additional observations.   Observations with  CHIRON started  in 2023.
They were  stopped for  administrative reasons  in 2023  September and
resumed  a  year later,  in  2024  August.   The pause  has  seriously
impacted      this      program,     precluding      optimum      time
coverage. Figure~\ref{fig:GKMorb} gives the RV curves deduced from the
CHIRON  spectra.  Individual  systems are  commented below,  and their
hierarchical  structure  is  illustrated  in  Figure~\ref{fig:mobile}.
There are five  quadruples of 3+1  architecture, one quintuple,
and one triple.

{\it GKM0} (CD$-$53 862) is revealed as an SB2 with a 120.7 day period
(Table~\ref{tab:sborb}) which matches the  Gaia astrometric orbit with
a period of 121.9 days. Gaia  did not determine the SB2 orbit possibly
because  of insufficient  spectral  resolution.  All  elements of  the
spectroscopic  orbit   are  fitted   without  restrictions.    The  RV
amplitudes and inclination  lead to the masses of 0.83  and 0.77 \msun
(mass ratio $q=0.93$).  The areas of  the CCF dips give the flux ratio
of $r=0.68$, so the $V$ magnitudes of spectroscopic components Aa1 and
Aa2  are  10.51  and  10.93  mag,  respectively,  and  their  absolute
magnitudes match the measured masses.  The full semimajor axis of this
orbit is  6.6\,mas, while the  axis of  the Gaia astrometric  orbit is
0.7\,mas. Owing to  the blending, the astrometric  amplitude should be
reduced by the factor  $f^* = q/(1+q) - r/(1+r) =  0.08$, close to the
actual ratio of 0.1.  If  blending is neglected, the astrometric orbit
gives the mass ratio of $\sim$0.15.   The faint ($\Delta I = 3.4$ mag)
speckle companion  at 0\farcs176 has  an estimated period $P^*$  of 38
yr, and  some orbital  motion is  detected in two  years; its  mass is
comparable  to the  masses of  Aa1 and  Aa2.  The  CPM companion  B at
36\farcs6 has an  estimated mass of 0.24 \msun. This  is a genuine 3+1
quadruple.

{\it GKM1} has a variable RV,  and the 4 available CHIRON observations
match the  36.6 day Gaia  SB1 orbit (I  fixed its elements  and fitted
only the  RV amplitude and the  systemic velocity). The Gaia  orbit is
plotted by dashed line in  Figure~\ref{fig:GKMorb}.  With the Aa1 mass
of 0.71 \msun, the minimum mass  of the spectroscopic secondary Aa2 is
0.21 \msun.  The speckle companion Ab at 0\farcs36 and 2.3 mag fainter
than Aa ($P^* \sim 154$ yr)  is confirmed indirectly by the multi-peak
transits of  star A in Gaia.   Two more Gaia  stars B and C,  making a
1\farcs3 pair,  are located at 153\arcsec  ~from A and have  common PM
and parallax.  The estimated masses of B and C, 0.7 and 0.4 \msun, are
comparable to the mass of Aa1.  The projected separation between A and
BC  is  14  kau,  and  the estimated  outer  period  is  $\sim$1  Myr.
Therefore, this  a 3+2 quintuple system  with a total mass  around 2.6
\msun.

% moving group?
%2022.1947   27.55  0.3636    2.29 q 2  I Nothinhg else!
% U,V,W:   -20.88  -25.02   -8.98

{\it GKM2} (LP 902-100)  has an SB2 orbit in the NSS  with a period of
37.17  days,  amplitudes of  37.55  and  35.66  \kms, and  a  systemic
velocity  of 96.4  \kms.  Only  single lines  are seen  in the  CHIRON
spectra, however.  The RV is  definitely variable and corresponds to a
quasi-circular orbit  with a period of  11.5 days and an  amplitude of
31.5 \kms. Both the amplitude and the systemic velocity are similar to
their values in the GDR3 orbit, while the periods differ dramatically.
However, the  GDR3 period aliased  by the  63 day Gaia  scanning cycle
corresponds to twice  the actual period: 1/37.17 + 1/63  = 1/23.4. One
might suspect that the Gaia pipeline for spectroscopic orbits measured
only single RVs in each scan but, somehow, alternatively assigned them
to two spectroscopic components, thus leading  to a false orbit with 
double   (and,  furthermore,   aliased)  period   and  comparable   RV
amplitudes.
The  speckle companion  has been  discovered in  2021.9 at  0\farcs139
separation ($P^* \sim 13$ yr), and  its fast motion is indeed detected
by two  subsequent observations in  2024 and 2025 (the  position angle
has  changed  by 77\degr).   The  magnitude  difference is  $\Delta  I
\approx 2.9$  mag, so  the companion  cannot explain  the double-lined
Gaia orbit.  The faint CPM star  B at 29\arcsec ~has estimated mass of
0.08 \msun and a spectral type M7.5 (DEA 21 pair in the WDS).  So,
this is a 3+1 quadruple where the Gaia SB2 orbit is spurious.

% sub-dwarf?

% 2021.8887  165.46  0.1386    2.87   2  I
% 2024.2383   99.39  0.0755    2.61 : 2  I
% 2025.0427   88.81  0.0299    2.41   2  I
% Gaia: V0=96.4 PM [-305,+151] CHIRON: V0=83.2
%U,V,W:   -82,4  -70.7    3.5

{\it  GKM3}  (TYC 8304-794-1)  has  a  constant  RV  of 31.6  \kms  (3
measurements).  The  RV amplitude  of the  Gaia 45  day orbit  is 10.0
\kms, the systemic  RV is 24.7 \kms;  this orbit is proven  here to be
spurious.  The  speckle companion at  0\farcs045 ($\Delta I  \sim 0.8$
mag) has a period  on the order of 5 yr; this pair  has closed down in
2022--2023 and was unresolved in  2024.18.  If its orbit is eccentric,
a  large  RV  variation  could   possibly  be  detected  by  Gaia  and
interpreted with a wrong period.  Star  B is at 15\farcs6. This system
is only a triple with component's masses between 0.6 and 0.7 \msun.

%  2022.1980   56.30  0.0445    1.35   2  I
%  2022.6844   50.03  0.0463    0.84 : 2  I
%  2023.1794   45.54  0.0414    1.16   2  I
%  2024.1569    0.00  0.0000    0.00 : 2  I
% DR3 V0=24.7 

{\it  GKM4} (CD$-$73  1491, $V=9.41$ mag)  is the  brightest target  in this
group.  The  306 day Gaia  orbit is  confirmed by CHIRON,  despite its
small amplitude of  2.2 \kms.  The CHIRON orbit fitted  to 7 RVs (with
the fixed Gaia  period) has a larger eccentricity and  an amplitude of
1.8 \kms,  with a  roughly matching systemic  velocity.  The  Gaia SB1
orbit, plotted  by dashed  line in Figure~\ref{fig:GKMorb},  is likely
biased by the scan-dependent RV  errors with an estimated amplitude of
1.4  \kms  caused  by  the  speckle  companion;  the  orbit  would  be
considered  spurious  if it  were  not  confirmed.  The  spectroscopic
secondary Aa2  has a  minimum mass  of 0.05  \msun, in  the substellar
domain.  The speckle companion at 0\farcs253 with $\Delta I = 3.5$ mag
has a period of $\sim$75 yr (its motion is detected) and the estimated
mass of  0.6 \msun.  The  CPM companion B at  19\farcs7 has a  mass of
0.16 \msun.  So, the solar-mass star Aa1 has three low-mass companions
arranged in a 3+1 hierarchy.

% 2022.3076   19.14  0.2532    3.50 q 2  I
% 2024.4661   27.02  0.2447    3.43 q 2  I
% V0=41.3
% q=0.06 f = 0.06 axis 10.6mas wobble 0.60 mas 

{\it GKM5:} The  four CHIRON RVs match the 28.55  day Gaia orbit (only
the systemic velocity $\gamma$ was adjusted by 1.65 \kms). The minimum
mass of  the spectroscopic  secondary Ab is  0.11 \msun.   The speckle
companion is  at 0\farcs09 ($P^*  \sim 16$  yr, the orbital  motion is
detected), and  the CPM  companion B  is at 4\farcs25.  This is  a 3+1
quadruple composed of K- and M-dwarf stars. 

% 2022.4415  323.77  0.0880    2.55 : 2  I
% 2023.5708  332.06  0.0757    2.43   2  I
% 2024.4661    0.00  0.0000    0.00 : 2  I
% V0=10.6

{\it GKM6} (HIP  58691, HD~104532) was observed 7 times,  and the RVs
match the  GDR3 orbit.  I fitted  the elements  $T$, $K_1$,  and $V_0$
while fixing the remaining elements to the respective GDR3 values. The
rms residuals are 0.09 \kms.

%---------------------------------------------------------
\section{Discussion}
\label{sec:sum}

The Gaia mission was designed to  deal with single stars and binaries.
Hierarchical systems are obvious  troublemakers for the automated Gaia
pipelines that  cannot take into  account such aspects as  spatial and
spectral blending  between components, the cross-talk  between spatial
and  spectral domains,  and multiple  periods. The  Gaia scanning  law
limits  the  range  of  accessible periods  and  the  phase  coverage,
especially for very short periods  or very eccentric orbits.  Although
clever  ways to  circumvent these  problems might  be invented  in the
future,  the need  of complementary  data on  hierarchical systems  to
pinpoint their true architecture is obvious.

The  four  solar-type hierarchies  studied  here,  as well  as  other
targets of this project \citep{chiron10}, illustrate the complexity and
diversity  of  situations, unlikely  to  be  handled properly  by  the
automatic  pipelines.   On  the  other  hand,  the  immense  discovery
potential of  Gaia is evident,  especially for stars brighter  than $G
\sim 13.5$ mag that have the RV  data.  An RV variability in excess of
the  orbital  motion expected  in  the  astrometric binaries  revealed
hundreds  of  new candidate  hierarchies  whose  inner periods  remain
unknown \citep{Bashi2024}.  This resembles the pre-Gaia situation when
the  survey of  \citet{N04}  discovered new  hierarchies, but  further
dedicated monitoring was needed to establish their architecture.

Among the  seven stars with  NSS orbits and speckle  companions tested
here, six are   3+1 (or 3+2) hierarchies, while GKM3
is only a triple.  The NSS orbits  of GKM2 and GKM3 are not confirmed.
The  results of  this  study  are incorporated  in  the multiple  star
catalog                                         \citep{MSC}.\footnote{
  \url{http://vizier.u-strasbg.fr/viz-bin/VizieR-4?-source=J/ApJS/235/6}
  and \url{https://www.ctio.noirlab.edu/~atokovin/stars/} }

%\acknowledgements
\begin{acknowledgments} 

I  thank  the   operators  of  the  1.5  m   telescope  for  executing
observations of  this program and  the SMARTS team for  scheduling and
pipeline processing.   T.~Johns has kindly  shared the spectra  of HIP
55691  taken in  2023 with  CHIRON  and used  here in  the orbit  fit.
Comments by D.~Bashi  and B.~Mason on the draft version  of this paper
are greatly appreciated.

The research  was funded  by the  NSF's NOIRLab.   This work  used the
SIMBAD   service  operated   by   Centre   des  Donn\'ees   Stellaires
(Strasbourg, France),  bibliographic references from  the Astrophysics
Data System  maintained by  SAO/NASA, and  the Washington  Double Star
Catalog maintained at  USNO.  This work has made use  of data from the
European       Space       Agency       (ESA)       mission       Gaia
(\url{https://www.cosmos.esa.int/gaia}),  processed by  the Gaia  Data
Processing        and         Analysis        Consortium        (DPAC,
\url{https://www.cosmos.esa.int/web/gaia/dpac/consortium}).    Funding
for the DPAC has been provided by national institutions, in particular
the  institutions participating  in the  Gaia Multilateral  Agreement.
This paper  includes data  collected with  the TESS  mission, obtained
from the  MAST data archive  at the Space Telescope  Science Institute
(STScI). Funding for the TESS mission is provided by the NASA Explorer
Program.  STScI is  operated by  the Association  of Universities  for
Research in Astronomy, Inc., under NASA contract NAS 5–26555.

%The TESS data were retrieved from the Barbara A. Mikulski Archive for Space Telescopes (MAST). 

%ESO archive.
\end{acknowledgments} 

\facility{CTIO:1.5m, SOAR, Gaia}

%---------------------------------------------------------
%\section{}
%\label{sec:}

%---------------------------------------------------------
%\section{}
%\label{sec:}

%---------------------------------------------------------
%\section{}
%\label{sec:}

%\subsection{}

%\input{paper9.bbl}

\bibliography{chiron.bib}
\bibliographystyle{aasjournal}

\end{document}